\documentclass[12pt]{iopart}
\usepackage{natbib}
\usepackage{epsfig}
\usepackage{iopams}
\graphicspath{{/home/cesko/Documents/Lavori/SemiclHam/Images/}}
\begin{document}
  \title[Mean-field algebraic approach]
{Mean-field algebraic approach to the dynamics of fermions in 
a 1D optical lattice}
  \author{F. P. Massel, V. Penna}
  \address{Dipartimento di Fisica, Torino Politecnico, Corso Duca degli
           Abruzzi 24, I-10129 Torino, Italy}
  \ead{francesco.massel@polito.it}
  \begin{abstract}
We consider a one-dimensional optical lattice of three-dimensional Harmonic
    Oscillators which are loaded with neutral fermionic atoms trapped into two hyperfine states.
    By means of a standard variational coherent-state procedure,
    we derive an effective Hamiltonian for this quantum model and
    the hamiltonian equations describing its evolution.
To this end, we identify the algebra $\mathcal L$ of two-fermion operators 
--describing the relevant microscopic quantum processes of our model-- whereby the natural choice for the trial state appears to be a so(2r) coherent state.
The coherent-state parameters, playing the role of dynamical variables
for the effective Hamiltonian, are shown to identify
with the $\mathcal L$-operator expectation values thus
providing a clear physical interpretation of this algebraic mean-field
picture.
  \end{abstract}  
  \submitto{\jpb}
  \maketitle
 
 \section{Introduction}
 Recently, a hierarchy of Hubbard-like Hamiltonians has been proposed to
 describe the behavior of ultracold fermions in one-dimensional optical
 lattices \cite{Massel}. 
 
\begin{figure}[htbp]
       \centering \epsfig{figure=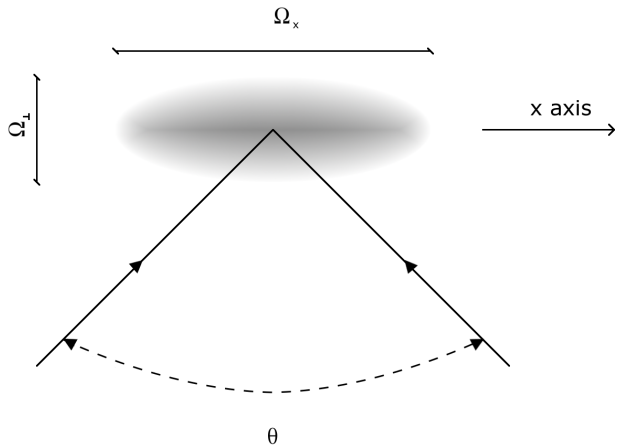,width=0.6\textwidth}
       \caption{Sketch of the experimental setup considered. Two laser beams,
         propagating at an angle $\theta$, give rise to an AC Stark induced
         periodic potential. In addition the fermionic cloud is confined by a
         cigar-shaped magnetic trap with trapping frequencies $\Omega_\perp$
         and $\Omega_x$.}
         \label{fig:Setup}
  \end{figure}

 These lattices can be realized with a pair of lasers propagating at a given
 angle $\theta$ ($\theta=\pi$ represents the familiar counterpropagating
 case), with global confinement ensured by a magnetic trap (see Fig.
 \ref{fig:Setup}, for a detailed description of this setup see \cite{Peil}).
 The pair of lasers give rise to a interference pattern needed to obtain a
 periodic potential by AC Stark effect. The lattice constant can be adjusted
 tuning the angle $\theta$ according to the relation $d=\lambda/2\sin
 (\theta/2)$ where $\lambda$ is the laser wavelength. In addition, it is
 possible to control both the barrier height of the periodic potential (as a
 function of the laser intensity) and the interaction between fermions via an
 external magnetic field (Feshbach resonance, see, e.g. \cite{Kokkelmans}).
 
 These simple considerations allow one to understand how ultracold-atoms physics
 offers the possibility to explore experimentally a wide range of parameters
 set that would be unattainable in other contexts, such as the Hubbard model
 in condensed matter physics.
   
As a first step towards the description of the cited models, we propose
here a mean-field algebraic approach based on coherent-states procedure
\cite{Gilmore} for a fermionic one-dimensional array of harmonic wells.  Although the analytical
 approach followed here may be regarded as completely general, future
 numerical analysis will concentrate on a dimer with a six-level structure per well as depicted in Fig.  \ref{fig:levels}). The approach followed here
allows one a straightforward reformulation of the
the usual mean-field approach for quantum system (based on the `linearization' of the Hamiltonian and the subsequent solving appropriate self-consistency equations) in terms of a corresponding classical dynamical system. While,
 for fermions, the interpretation of the aforementioned classical dynamical
 system as a semiclassical approximation seems not beyond need of
 justification, it is possible to give a precise physical interpretation to
 the dynamical variables of the classical problem in terms of expectation
 values of quantum operators.
 
 In general, it is possible to consider a mean-field approach to a given
 problem as the constrained minimization of the Hamiltonian $\hat{H}$ over
 a algebra $\mathcal{L}$. A different choice of $\mathcal{L}$ will lead to
 different mean-field solutions (\cite{Rasetti,Gilmore,MontoPe}).
 In particular, 
 we will focus on the so(2r) coherent states that, as it will
 be shown, will lead to the Hartree-Fock-Bogoliubov \cite{Lieb} mean-field
 approximation, whose effectiveness has been proven for a single spherical
 harmonic trap in \cite{Grasso}.
 
The paper is organized as follows.
 In section \ref{sec:ModDesc} a brief discussion of the general model
 considered will be given, along with some possible approximations in
 different physical situations. As we already mentioned, the 
 fully-analytical control over the physical parameters allows to conceive various Hamiltonians that may have direct experimental relevance.  
In section \ref{sec:CohSt} so(2r) coherent states and the relevant algebra
 will be defined. The end of this rather technical section will be devoted to
 the physical interpretation of the choice of so(2r) as the algebra for the
 mean-field procedure.
 In section \ref{sec:ClH} the classical Hamiltonian $\mathcal{H}_{cl}$ will
 be deduced and the functional dependence in terms of quantum operators
 expectation values will be investigated.
Finally, in section \ref{sec:EvZeta} the analysis of the classical dynamical system
 whose Hamiltonian is $\mathcal{H}_{cl}$ is performed: Lie-Poisson brackets
(namely the `classical' commutators) and, consequently, the evolution equations for the dynamical variables are given.

\section{Model Description}
  \label{sec:ModDesc}
 
 In \cite{Massel}, along the lines introduced in \cite{Albus}, 
 a generalized Hubbard Hamiltonian has been introduced to
 describe the behavior of alkali-metal fermionic atoms in a one-dimensional
 optical lattice of oblate three dimensional (2+1D) Harmonic Oscillators (pancakes) 
 \begin{equation}
  \label{eq:RHH}
  \hat{\mathsf{H}}= \sum_{\alpha} 
                                   \lambda_{\alpha}
                                 \hat{n}_\alpha+ 
                              \sum_{\alpha,\beta} 
                                    T_{\alpha,\beta} 
                                   \hat{c}^\dagger_\alpha
                                   \hat{c}_\beta
                       +\sum_{\alpha,\beta,\gamma,\delta} 
                                     U_{\alpha,\beta,\gamma,\delta}
                                   \hat{c}^\dagger_\alpha
                                   \hat{c}^\dagger_\beta
                                   \hat{c}_\delta
                                   \hat{c}_\gamma \, .  
\end{equation}

In Eq. (\ref{eq:RHH}), $\alpha$ must be considered as a multiple index
$\alpha=\left\{i_\alpha,n_\alpha=0,J_ \alpha,m_\alpha,\sigma_\alpha \right\}$
whose origin can be traced back to the space(+local)-modes approximation.  In
this picture $n_\alpha,J_\alpha$ and $m_\alpha$ are the local 2+1 D Harmonic
Oscillator quantum numbers, $i_\alpha$ is the site quantum number and
$\sigma_\alpha$ is the spin quantum number. In the following we will confine
our analysis to situation where radial modes only are involved in the system
dynamics (i.e. we will ``freeze'' the axial quantum number $n_\alpha$ to
zero). The validity of this assumption is guaranteed as long as the radial
trapping frequency $\Omega_\perp$ is much smaller than the axial trapping
frequency $\omega_x$, i.e. $\Omega_\perp/\omega_x \ll 1$. In this case the tunneling coefficient assumes the following
form:
\begin{equation}
      \label{eq:Tdelta}
      T_{\alpha,\beta}=\delta_{J_\alpha,J_\beta}
                       \delta_{m_\alpha,m_\beta} 
                       \delta_{\sigma_\alpha,\sigma_\beta}
                       T_{i_\alpha,i_\beta}
\end{equation}
if we allow nearest-neighbor hopping only
\begin{equation}
  \label{eq:Tdelta2}
   T_{i_\alpha,i_\beta}=T\delta_{i_\alpha+1,i_\beta} \, ,
\end{equation}
where $T$ is a known function of the external parameters.  Another assumption
concerns the two-body interaction term $U_{\alpha,\beta,\gamma,\delta}$, which
is treated within the pseudopotential approximation, leading to a delta-like spatial
dependence, thus excluding nearest-neighbor two-body interaction terms. If the
fermionic nature of the interacting particles is taken into account we have
\begin{equation}
  \label{eq:U}
   U_{\alpha,\beta,\gamma,\delta}=
   \delta_{i_\alpha,i_\beta,i_\gamma,i_\delta} 
   \delta_{\sigma_\alpha,\sigma_\gamma}\delta_{\sigma_\beta,\sigma_\delta}
\delta_{\sigma_\alpha,-\sigma_\beta}
U_{\{J_\alpha,m_\alpha\},\{J_\beta,m_\beta\},\{J_\gamma,m_\gamma\},\{J_\delta,m_\delta\}}\,.
\end{equation}
Finally, we give the expression for the one-particle energy term which is
essentially given by the single-particle energy of the 3D Harmonic Oscillator
\begin{equation}
      \label{eq:lambda2}
       \lambda_{i_\alpha,n_\alpha,J_\alpha,m_\alpha,\sigma_\alpha}= 
          \left[
                \hbar\omega_x
              \left(
                  n_\alpha+\frac{1}{2}
              \right)+  
                \hbar\Omega_\perp\left(2J_\alpha+1\right) - T_{i_\alpha,i_\alpha}
          \right] \, ,
\end{equation}
where $T_{i_\alpha,i_\alpha}$ represents a ``hopping correction'' to the
single particle energy term.
%
%In this paper we will consider a system constituted by a chain whose
%sites with six
%single-particle states each see Fig.\ref{fig:levels}.

\begin{figure}[htbp]
       \centering \epsfig{figure=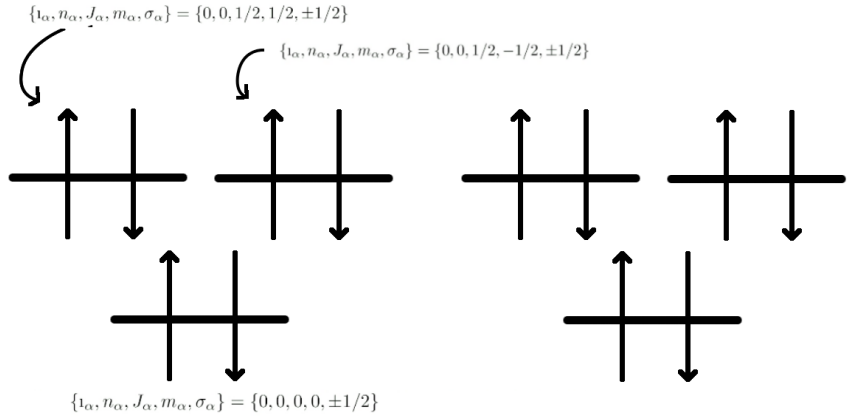,width=0.8\textwidth}
       \caption{Schematic view of the single-particle energy levels. Up and
         down arrows represent possible spin values. We have explicitly
         indicated single-particle quantum numbers for the left-hand-side well.
         Analogously, it is possible to define quantum numbers for the
         right-hand-side well.}  \label{fig:levels}
 \end{figure}

For the case considered in Fig. \ref{fig:levels} the selection rules imposed
on the two-body interaction term select three possible values on
$U_{\alpha,\beta,\gamma,\delta}$ that can be classified as: i) lowest-level/
lowest-level interaction terms, ii) lowest-level/highest-level interaction
terms, iii) highest-level/highest-level interaction terms.

\section{Coherent States}
  \label{sec:CohSt}
  Since in our mean-field analysis we would like to keep trace of the
  (possible) atom pairing, the most appropriate coherent-states algebra seems
  to be, according to \cite{Gilmore}, the algebra spanned by the $r(2r-1)$
  operators $\{\hat{c}_\alpha^\dagger\hat{c}_\beta (1\leq i,j\leq r),
  \hat{c}_\alpha\hat{c}_\beta,\hat{c}^\dagger_\alpha\hat{c}^\dagger_\beta \}$,
  i.e. so(2r). Its commutation relations can be written as
  \begin{eqnarray}
   \label{eq:CommY}
   &&\left[ Y^1_{\alpha\beta},Y^1_{\gamma\delta}\right]=\left[
     Y^2_{\alpha\beta},Y^2_{\gamma\delta}\right]=0 \nonumber\\
   &&\left[ Y^1_{\alpha\beta},Y^2_{\gamma\delta}\right]=
     Y^3_{\gamma\beta}\delta_{\alpha\delta}+Y^3_{\delta\alpha}\delta_{\beta\gamma}
     -Y^3_{\gamma\alpha}\delta_{\beta\delta}-Y^3_{\delta\beta}\delta_{\alpha\gamma}\nonumber\\
   &&\left[Y^1_{\alpha\beta},Y^3_{\gamma\delta}\right]=
     Y^1_{\alpha\delta}\delta_{\beta\gamma}-Y^1_{\beta\delta}\delta_{\alpha\gamma}\nonumber\\
   &&\left[Y^2_{\alpha\beta},Y^3_{\gamma\delta}\right]=Y^2_{\beta\gamma}\delta_{\alpha\delta}-Y^2_{\alpha\gamma}\delta_{\beta\delta}\, ,
  \end{eqnarray}
  having defined
  \begin{eqnarray}
    \label{eq:AlgComm}
     Y^1_{\alpha\beta}=\hat{c}_\alpha\hat{c}_\beta\, ,\\
     Y^2_{\alpha\beta}=\hat{c}^\dagger_\beta\hat{c}^\dagger_\alpha\, , \\
     Y^3_{\alpha\beta}=\hat{c}^\dagger_\alpha\hat{c}_\beta \,.
  \end{eqnarray}
   With the above definitions the coherent states can be expressed as
  \begin{equation}
    \label{eq:CohSt}
      |\phi\rangle=\exp 
\left [ \, - \sum_{1\leq \alpha \neq \beta \leq r}
            \left( \eta_{\alpha,\beta}\hat{c}^\dagger_\alpha
        \hat{c}^\dagger_\beta-H.c.\right) \right ] |0>\,.             
  \end{equation}
  
  To evaluate the expectation value of the Hamiltonian $\hat{H}$ defined by
  equation (\ref{eq:RHH}) over the coherent state of the form (\ref{eq:CohSt}),
  it is necessary to evaluate the action of the operator
  \begin{equation}
     \label{eq:Omega}
       \hat{\Omega} = 
\exp 
\left [ \, -\sum_{1\leq \alpha \neq \beta \leq r}
            \left( \eta_{\alpha,\beta}\hat{c}^\dagger_\alpha
        \hat{c}^\dagger_\beta-H.c.\right) \right ]      
    \end{equation}    
  over the fermionic raising and lowering operators. Namely
  \begin{eqnarray}
   \label{eq:ExplOmOp}
   \hat{\Omega}^\dagger \hat{c}^\dagger_\alpha \hat{\Omega} =
      \exp &&\left[
             \sum_{1\leq i \neq j \leq r}
            \left( 
              \eta_{i,j}
                  \hat{c}^\dagger_i \hat{c}^\dagger_j - 
              \eta^*_{i,j}
                  \hat{c}_j\hat{c}_i 
            \right)
           \right] 
         \hat{c}^\dagger_\alpha  \nonumber \\
      &&\exp \left[
             -\sum_{1\leq i \neq j \leq r}
            \left( 
              \eta_{i,j}
                  \hat{c}^\dagger_i \hat{c}^\dagger_j - 
              \eta^*_{i,j}
                  \hat{c}_j\hat{c}_i 
            \right)
           \right]
 \end{eqnarray}
which, exploiting the BCH formula \cite{Gilmore} can be written as
\begin{equation}
   \label{eq:OpBCH}
  \hat{\Omega}^\dagger \hat{c}^\dagger_\alpha \hat{\Omega} = 
          \sum_m \frac{1}{m} \left[ \sum_{ij} \eta_{i,j}
                  \hat{c}^\dagger_i \hat{c}^\dagger_j - 
              \eta^*_{i,j}
                  \hat{c}_j\hat{c}_i \right]_m \,.
 \end{equation}
 It can be shown that in the last summation the two first terms only survive,
 leading to to the following expression for $\hat{\Omega} 
 \hat{c}^\dagger_\alpha \hat{\Omega}^\dagger$ and $\hat{\Omega} \hat{c}_\gamma
 \hat{\Omega}^\dagger$ respectively
 \begin{eqnarray*}
   \label{eq:OmExpl4}
   &&\hat{\Omega}^\dagger \hat{c}^\dagger_\alpha \hat{\Omega}=
      \hat{c}^\dagger_\alpha +\sum_i \zeta^*_{i\alpha}\hat{c}_i \, ,\\
   &&\hat{\Omega}^\dagger \hat{c}_\gamma \hat{\Omega}=
   \hat{c}_\gamma + \sum_i\zeta_{m\gamma} \hat{c}^\dagger_m
 \end{eqnarray*}
 with $\zeta_{ij}=2\eta_{ij}$.

 We are now in the position to evaluate 
 \begin{equation}
   \label{eq:ExpVal}
   \mathcal{H}_{cl}=\langle\phi|\hat{\mathsf{H}}-\mu \hat{N}|\phi\rangle\, ,
 \end{equation}
 where the term $\mu \hat{N}$ has been added to take into account the particle
 number constraint.

 With Eq. (\ref{eq:CohSt}), Eq. (\ref{eq:ExpVal}) becomes
 \begin{equation}
   \label{eq:ExpVal2}
    \mathcal{H}_{cl}=\langle0|\Omega\left[\left(\hat{\mathsf{H}}_0+\hat{\mathsf{H}}_I\right)\right]\Omega^\dagger|0\rangle
    \, ,
 \end{equation}
 where 
 \begin{eqnarray*}
   \hat{\mathsf{H}}_0= \sum_{\alpha,\beta} 
                         \Gamma_{\alpha,\beta} 
                           \hat{c}^\dagger_\alpha
                           \hat{c}_\beta \, , \\
   \hat{H}_I=\sum_{\alpha,\beta,\gamma,\delta} 
                                     U_{\alpha,\beta,\gamma,\delta}
                                   \hat{c}^\dagger_\alpha
                                   \hat{c}^\dagger_\beta
                                   \hat{c}_\delta
                                   \hat{c}_\gamma \, ,
 \end{eqnarray*}
 with $\Gamma_{\alpha,\beta}=\lambda_\alpha\delta_{\alpha,\beta}-T_{\alpha,\beta}-\mu \delta_{\alpha,\beta}.$
 Since $\Omega$ is a unitary operator, we can write
 \begin{eqnarray}
   \label{eq:OmH}
      \hat{\Omega}^\dagger \hat{H}_0 \hat{\Omega} =
                   \sum_{\alpha,\beta} 
                         \Gamma_{\alpha,\beta} 
                         \hat{\Omega}^\dagger 
                           \hat{c}^\dagger_\alpha
                         \hat{\Omega} 
                         \hat{\Omega}^\dagger 
                           \hat{c}_\beta
                         \hat{\Omega}\, , \nonumber \\  
      \hat{\Omega}^\dagger \hat{H}_I \hat{\Omega} =
                   \sum_{\alpha,\beta,\gamma,\delta} 
                                     U_{\alpha,\beta,\gamma,\delta}
                               \hat{\Omega}^\dagger     \hat{c}^\dagger_\alpha \hat{\Omega} 
                               \hat{\Omega}^\dagger     \hat{c}^\dagger_\beta  \hat{\Omega} 
                               \hat{\Omega}^\dagger     \hat{c}_\delta         \hat{\Omega} 
                               \hat{\Omega}^\dagger     \hat{c}_\gamma         \hat{\Omega} \, . 
 \end{eqnarray}

 The following expectation values must then be evaluated. For the one-body term
 \begin{equation}
   \label{cs2}
     \hat{\Omega}^\dagger \hat{H}_0 \hat{\Omega} =
                   \sum_{\alpha,\beta} 
                         \Gamma_{\alpha,\beta} 
                         \left[
                           \hat{c}^\dagger_\alpha +\sum_i \zeta^*_{i\alpha} \hat{c}_i
                         \right] \cdot  
                         \left[
                           \hat{c}_\beta + \sum_k \zeta_{k\beta} \hat{c}^\dagger_k 
                         \right]
\end{equation}
and for the interaction term
  \begin{eqnarray}
   \label{cs2.1}
    \hat{\Omega}^\dagger \hat{H}_I \hat{\Omega} =&&\sum_{\alpha,\beta,\gamma,\delta} 
        U_{\alpha,\beta,\gamma,\delta}
         \left[
         \hat{c}^\dagger_\alpha +\sum_i \zeta^*_{i\alpha} \hat{c}_i
       \right] \cdot 
       \left[
         \hat{c}^\dagger_\beta +\sum_j \zeta^*_{j\beta} \hat{c}_j
       \right] \cdot \nonumber\\
       &&\hspace{-1cm} \left[
         \hat{c}_\delta + \sum_k \zeta_{k\delta} \hat{c}^\dagger_k 
       \right] \cdot
       \left[
         \hat{c}_\gamma + \sum_l \zeta_{l\gamma} \hat{c}^\dagger_l 
       \right]  \,.
 \end{eqnarray}
 
 As it can be directly verified, in the calculation of the expectation values
 over the vacuum state $|0\rangle$ only the following terms survive
 \begin{equation}
   \label{eq:cs3}
   \langle0|\hat{\Omega}^\dagger \hat{H}_0
   \hat{\Omega} |0\rangle=\sum_{\alpha,\beta} \sum_{ij} 
   \Gamma_{\alpha,\beta}\zeta^*_{i\alpha}\zeta_{j\beta}
   \langle0|\hat{c}_i\hat{c}^\dagger_j|0\rangle \, ,
   \end{equation}
   \begin{eqnarray}
    \label{eq:cs3.1} 
   &&\langle0|\hat{\Omega}^\dagger \hat{H}_I \hat{\Omega} |0\rangle= 
    \sum_{\alpha,\beta,\gamma,\delta} 
      U_{\alpha,\beta,\gamma,\delta} 
     \left(
       \sum_{i,j,k,l}
      \zeta^*_{i\alpha}\zeta^*_{j\beta}\zeta_{k\delta}\zeta_{l\gamma}
         \langle0| \hat{c}_i \hat{c}_j\hat{c}^\dagger_k
   \hat{c}^\dagger_l|0\rangle \right. \nonumber \\
   &&\hspace{5.5cm}+\left.
        \sum_{i,j}\zeta^*_{i\alpha}\zeta_{j\gamma}\langle0|\hat{c}_i\hat{c}^\dagger_\beta\hat{c}_\delta\hat{c}^\dagger_j|0 \rangle  
     \right)\,.
 \end{eqnarray}
The two expectation values over the vacuum state give 
 \begin{eqnarray}
   \label{eq:ExpVal3}
   && \langle0| \hat{c}_i \hat{c}^\dagger_j|0\rangle=\delta_{ij} \\
   &&\langle0| \hat{c}_i \hat{c}_j\hat{c}^\dagger_l \hat{c}^\dagger_k|0\rangle =
     \delta_{il}\delta_{jk}-\delta_{jk}\delta_{il} \\
   &&\langle0|\hat{c}_i\hat{c}^\dagger_\beta\hat{c}_\delta\hat{c}^\dagger_j|0
     \rangle=
     \delta_{i\beta}\delta_{j\delta}
 \end{eqnarray}
hence 
\begin{equation}
    \label{eq:Hcl}
\hspace{-2cm}
\mathcal{H}^{cl}=\sum_{\alpha\beta} \Gamma_{\alpha\beta}\sum_{i}
    \zeta^*_{i\alpha}\zeta_{i\beta} +
    \sum_{\alpha,\beta,\gamma,\delta} 
    U_{\alpha,\beta,\gamma,\delta} 
    \left[
      \sum_{i,j} \left( \zeta^*_{i\alpha}\zeta^*_{j\beta}\zeta_{j\delta}\zeta_{i\gamma}-\zeta^*_{i\alpha}\zeta^*_{j\beta}\zeta_{i\delta}\zeta_{j\gamma} \right) +
        \zeta^*_{\beta\alpha}\zeta_{\delta\gamma}  
    \right]
\end{equation}
 
\section{Effective Hamiltonian }
  \label{sec:ClH}

Hamiltonian (\ref{eq:Hcl}) can be shown to represent the effective Hamiltonian 
associated with $\hat H$ within the time-dependent variational principle
procedure \cite{Gilmore}. The latter is based on approximating
the quantum states of the system by a trial state
$|\Psi \rangle$ satisfying the weak form of the 
Schr\"odinger equation 
$ \langle \Psi |i \hbar \partial_t -{\hat H}| \Psi \rangle = 0 $.
Here, we assume that $| \Psi \rangle$, up to an irrelevant phase factor,
is the coherent state defined in equation (\ref{eq:CohSt}).
The variational procedure allows one to derive the effective Lagrangian
${\dot S} = \langle\Psi |i \hbar \partial_t -{\hat H}| \Psi\rangle $,
depending on dynamical variables $\zeta_{\alpha \beta}$, which in turn
supplies the effective Hamiltonian (\ref{eq:Hcl}).
Such a procedure provides as well the dynamical equations pertaining to
Hamiltonian (\ref{eq:Hcl}) and the relevant Lie-Poisson brackets. 
The latter exhibit the same algebraic structure of
commutators (\ref{eq:CommY}) and will be defined below.

A quite direct physical insight about coherent-state parameters 
$\zeta_{\alpha \beta}$ is achieved when considering 
the expectation values for the elements of the Lie
  algebra so(2r) over the coherent states $|\phi\rangle$. We have
  \begin{eqnarray}
   \label{eq:Exp1}
  &&\langle \phi|\hat{c}^\dagger_\alpha\hat{c}^\dagger_\beta |\phi
  \rangle=\zeta^*_{\beta\alpha} \, ,\\
  &&\langle \phi|\hat{c}^\dagger_\alpha\hat{c}_\beta |\phi \rangle=\sum_i
  \zeta^*_{i\alpha}\zeta_{i\beta}=\xi_{\alpha\beta} \, ,\\
  &&\langle \phi|\hat{c}_\alpha\hat{c}_\beta |\phi \rangle=\zeta_{\alpha\beta}
  \, ,
  \end{eqnarray}
showing how parameters $\zeta_{\alpha \beta}$ are related
to microscopic physical processes of creation/destruction of lattice fermions. 
Moreover Eq. (\ref{eq:Hcl}) can be
  written as
  \begin{equation}
    \label{eq:HClSpin}
    \mathcal{H}_{cl}=\sum_{\alpha\beta} \Gamma_{\alpha\beta} \xi_{\alpha\beta} +
    \sum_{\alpha,\beta,\gamma,\delta} 
       U_{\alpha,\beta,\gamma,\delta} 
    \left[
          \left( \xi_{\alpha\gamma}\xi_{\beta\delta}- 
                            \xi_{\alpha\delta}\xi_{\beta\gamma}
                     \right) +
          \zeta^*_{\beta\alpha}\zeta_{\delta\gamma} 
    \right]\, .
  \end{equation}
  The three terms represent the direct, the exchange and the pairing term in
  the HFB mean-field approximation.  

\subsection{Evolution Equations for the canonical variables}
  \label{sec:EvZeta}
  According to \cite{Gilmore} the variables $\zeta_{\alpha\beta}$,
  $\zeta^*_{\alpha\beta}$ and $\xi_{\alpha \beta}$ represent the canonical
  variable for the classical Hamiltonian $\mathcal{H}_{cl}$. With a well-known
  procedure \cite{Perelomov} it is possible to describe the time evolution of
  those canonical variables in terms of their Poisson brackets with
  $\mathcal{H}_{cl}$.
  To write the Lie-Poisson brackets for the given dynamical system we can make explicit
  the structure constants for the so(2r) algebra
  \begin{eqnarray}
   \label{eq:StrConst}
   &&c_{1,\alpha,\beta;1,\gamma,\delta}^{\phantom{1,\alpha,\beta;1,\gamma,\delta}q,\mu,\nu}=
     c_{2,\alpha,\beta;2,\gamma,\delta}^{\phantom{2,\alpha,\beta;2,\gamma,\delta}q,\mu,\nu}=0
     \nonumber \, ,\\
   &&c_{1,\alpha,\beta;2,\gamma,\delta}^{\phantom{1,\alpha,\beta;2,\gamma,\delta}q,\mu,\nu}=\delta_{q,3}\left(
                                            \delta_{\mu\gamma}\delta_{\nu\beta}\delta_{\alpha\delta}+\delta_{\mu\delta}\delta_{\nu\alpha}
                                            \delta_{\beta\gamma}-
                                            \delta_{\mu\gamma}\delta_{\nu\alpha}\delta_{\beta\delta}-\delta_{\mu\delta}\delta_{\nu\beta}
                                            \delta_{\alpha\gamma}
                                      \right) \nonumber \, ,\\
   &&c_{1,\alpha,\beta;3,\gamma,\delta}^{\phantom{1,\alpha,\beta;3,\gamma,\delta}q,m,n}=\delta_{q,1}\left(
                                            \delta_{\mu\alpha}\delta_{\nu\delta}\delta_{\beta\gamma}-
                                            \delta_{\mu\beta}\delta_{\nu\delta}\delta_{\alpha\gamma}
                                      \right) \nonumber \, ,\\                                    
  &&c_{2,\alpha,\beta;3,\gamma,\delta}^{\phantom{2,\alpha,\beta;3,\gamma,\delta}q,m,n}=\delta_{q,2}\left(
                                            \delta_{\mu\beta}\delta_{\nu\gamma}\delta_{\alpha\delta}-
                                            \delta_{\mu\alpha}\delta_{\nu\gamma}\delta_{\beta\delta}
                                      \right) \, .
 \end{eqnarray}

Thus the Poisson brackets have the following form
\vfill\eject
\begin{eqnarray}
  \label{eq:PoissBrack}
\hspace{-2cm}  
\left\{f,g \right\} = && \sum_{\alpha\beta\gamma\delta}
%\nonumber \\&&
\left(\xi_{\gamma\beta}\delta_{\alpha\gamma}-
        \xi_{\gamma\alpha}\delta_{\beta\delta}+
        \xi_{\delta\alpha}\delta_{\beta\gamma}-
        \xi_{\delta\beta}\delta_{\alpha\gamma}
  \right)
     \left(
           \frac{\partial f}{\partial \zeta_{\alpha\beta}}
           \frac{\partial g}{\partial \zeta^*_{\gamma\delta}}-
           \frac{\partial f}{\partial \zeta^*_{\gamma\delta}}
           \frac{\partial g}{\partial \zeta_{\alpha\beta}}
     \right) + \nonumber \\
    && +\left(
           \zeta_{\alpha\delta}\delta_{\gamma\beta}-\zeta_{\beta\delta}\delta_{\gamma\alpha}
     \right)
     \left(
           \frac{\partial f }{\partial \zeta_{\alpha\beta}} 
           \frac{\partial g }{\partial \xi_{\gamma\delta}}-
           \frac{\partial f }{\partial \xi_{\gamma\delta}}
           \frac{\partial g }{\partial \zeta_{\alpha\beta}}
     \right)+ \nonumber \\
     &&+ \left(
           \zeta_{\gamma\beta}\delta_{\alpha\delta}-\zeta_{\alpha\gamma}\delta_{\beta\delta}
     \right)
     \left(
           \frac{\partial f }{\partial \zeta^*_{\alpha\beta}} 
           \frac{\partial g }{\partial \xi_{\gamma\delta}}-
           \frac{\partial f }{\partial \xi_{\gamma\delta}}
           \frac{\partial g }{\partial \zeta^*_{\alpha\beta}}
     \right)+ \nonumber \\
     && +\left(
           \xi_{\beta\delta}\delta_{\gamma\alpha}-\xi_{\alpha\gamma}\delta_{\beta\delta}
     \right)
     \left(
           \frac{\partial f }{\partial \xi_{\alpha\beta}} 
           \frac{\partial g }{\partial \xi_{\gamma\delta}}-
           \frac{\partial f }{\partial \xi_{\gamma\delta}}
           \frac{\partial g }{\partial \xi_{\alpha\beta}}
     \right) \, .
\end{eqnarray}
Remembering that
\begin{equation}
  \label{eq:EvZeta}
  \dot{\zeta}_{\rho,\theta}=\left\{\zeta_{\rho,\theta},H\right\} \, ,
\end{equation}
it is possible to write
\begin{eqnarray}
  \label{eq:EvZeta2}
  \dot{\zeta}_{\rho,\theta}=&&
   \sum_\alpha
   \left(\Gamma_{\rho\alpha}\zeta_{\theta\alpha}-\Gamma_{\theta\alpha}\zeta_{\rho\alpha}\right)+ \nonumber
   \\ &&
   \sum_{\alpha\gamma\eta}\left[U_{\alpha\rho\gamma\eta}\left(\zeta_{\theta\gamma}\xi_{\alpha\eta}-\zeta_{\theta\eta}\xi_{\alpha\gamma}\right)
   +U_{\alpha\theta\gamma\eta}\left(\zeta_{\rho\eta}\xi_{\alpha\gamma}-\zeta_{\theta\gamma}\xi_{\alpha\eta}\right)\right]
   \nonumber \\ &&
   \sum_{\beta\gamma\eta}\left[U_{\rho\beta\gamma\eta}\left(\zeta_{\theta\eta}\xi_{\beta\gamma}-\zeta_{\theta\gamma}\xi_{\beta\eta}\right)
   +U_{\theta\beta\gamma\eta}\left(\zeta_{\rho\eta}\xi_{\beta\gamma}-\zeta_{\rho\gamma}\xi_{\beta\eta}\right)\right]+
   \nonumber \\&&
   \sum_{\alpha\gamma\eta}
   \zeta_{\eta\gamma}\left(U_{\alpha\theta\gamma\eta}\xi_{\alpha\rho}-U_{\alpha\rho\gamma\eta}\xi_{\alpha\theta}\right)+
   \nonumber \\&&
    \sum_{\beta\gamma\eta}
   \zeta_{\eta\gamma}\left(U_{\rho\beta\gamma\eta}\xi_{\beta\theta}-U_{\theta\rho\gamma\eta}\xi_{\beta\rho}\right)
\end{eqnarray}
that provide the set of dynamical equations governing the evolution of the
coherent state that approximates the system quantum state. In particular, they
allow to find the mean-field ground the state for the system and to perform a
weakly-excited state analysis.

\section{Conclusions}
\label{sec:Concl}
In this paper we have formulated an HFB mean-field approximation for a
one-dimensional array of oblate Harmonic Oscillators loaded with neutral fermionic
atoms. As already pointed out by Grasso et al. \cite{Grasso}, the numerical
solution to Eq. (\ref{eq:EvZeta2}) appears to be rather demanding from a
computational point of view. It seems then appropriate for future work to
concentrate on the simplest situation beyond known models like, as already
mentioned, a dimer with a six-level local structure.

In this case the evaluation of ground-state properties in this mean-field
picture as a function of the relevant parameters(i.e $T_{\alpha,\beta}$,
$U_{\alpha\beta\gamma\delta}$, $\mu$) reduces to the fixed-point analysis of
Eq.(\ref{eq:EvZeta2}). Moreover, an extension to finite-temperature properties does not seem beyond the possibilities of the analytical techniques here outlined and may represent one of the future lines of research.

\appendix

\section{Expectation values calulation}

In the present section we will explicitly calculate the terms obtained form
Eqs. (\ref{cs2},\ref{cs2.1}) leading to Eqs. (\ref{eq:cs3}). To evaluate
(\ref{cs2}) we need to perform the following product calculation
\begin{equation}
  \label{eq:app1}
  \hat{\Omega}^\dagger \hat{H}_0
   \hat{\Omega} = \sum_{\alpha\beta ij} 
   \Gamma_{\alpha,\beta} \left[
         \hat{c}^\dagger_\alpha +\sum_i \zeta^*_{i\alpha}\hat{c}_i
       \right] \cdot
       \left[
         \hat{c}_\beta +\sum_j \zeta_{j\beta}\hat{c}^\dagger_j
       \right]
\end{equation}
which is equal to

\begin{equation}
  \label{eq:app2}
   \hat{\Omega}^\dagger \hat{H}_0
   \hat{\Omega} = \sum_{\alpha\beta}\left[\hat{c}^\dagger_\alpha \hat{c}_\beta + \sum_i \zeta^*_{i\alpha}\hat{c}_i
   \hat{c}_\beta+ 
   \sum_j \zeta_{j\beta} \hat{c}^\dagger_\alpha\hat{c}^\dagger_j +\sum_{ij}
   \zeta^*_{i\alpha}\zeta_{j\beta}\hat{c}_i \hat{c}^\dagger_j\right].
\end{equation}
The evaluation of Eq. (\ref{eq:app2}) over the vacuum state $|0\rangle$ leads
to vanishing contributions for all non number-conseving terms and for all the
terms with a lowering operator on the right-hand side (or a raising operator
on the left-hand side). Namely (see Eq. (\ref{eq:cs3})
 \begin{equation}
  \label{eq:app4}
  \langle0|\hat{\Omega}^\dagger \hat{H}_0
   \hat{\Omega} |0\rangle=\sum_{\alpha\beta} \sum_{ij} 
   \Gamma_{\alpha,\beta} \zeta^*_{i\alpha}\zeta_{j\beta} \langle0|\hat{c}_i\hat{c}^\dagger_j|0\rangle. 
 \end{equation}
 
 With an analogous procedure it is possible to evaluate the expression given
 by Eq. (\ref{cs2.1})
 \begin{eqnarray}
   \label{eq:app5}
    \hat{\Omega}^\dagger \hat{H}_I \hat{\Omega} =\sum_{\alpha,\beta,\gamma,\delta} 
        U_{\alpha,\beta,\gamma,\delta}
         \left[
         \hat{c}^\dagger_\alpha +\sum_i \zeta^*_{i\alpha} \hat{c}_i
       \right]&& \cdot 
       \left[
         \hat{c}^\dagger_\beta +\sum_j \zeta^*_{j\beta} \hat{c}_j
       \right] \cdot \nonumber\\
       &&\hspace{-1cm} \left[
         \hat{c}_\delta + \sum_k \zeta_{k\delta} \hat{c}^\dagger_k 
       \right] \cdot
       \left[
         \hat{c}_\gamma + \sum_l \zeta_{l\gamma} \hat{c}^\dagger_l 
       \right]  
 \end{eqnarray}
leading to
\begin{eqnarray}
  \label{eq:app6}
   && \hat{\Omega}^\dagger \hat{H}_I \hat{\Omega} =
      \sum_{\alpha,\beta,\gamma,\delta} 
        U_{\alpha,\beta,\gamma,\delta} \nonumber \\
  &&\left[\hat{c}^\dagger_\alpha \hat{c}^\dagger_\beta 
     \hat{c}_\delta \hat{c}_\gamma +
  \sum_k \zeta_{k\delta} 
     \hat{c}^\dagger_\alpha \hat{c}^\dagger_\beta 
      \hat{c}^\dagger_k \hat{c}_\gamma+
  \sum_{l} \zeta_{l\gamma}
     \hat{c}^\dagger_\alpha \hat{c}^\dagger_\beta 
     \hat{c}_\delta \hat{c}^\dagger_l+ 
  \sum_{kl} \zeta_{l\gamma} \zeta_{k\delta} 
     \hat{c}^\dagger_\alpha \hat{c}^\dagger_\beta 
     \hat{c}^\dagger_k \hat{c}^\dagger_l+ \right. \nonumber\\
 && \sum_j \zeta^*_{j\beta}  
     \hat{c}^\dagger_\alpha \hat{c}_j 
     \hat{c}_\delta \hat{c}_\gamma + 
  \sum_{jk} \zeta^*_{j\beta} \zeta_{k\delta} 
     \hat{c}^\dagger_\alpha \hat{c}_j 
     \hat{c}^\dagger_k \hat{c}_\gamma + 
  \sum_{jl} \zeta^*_{j\beta}\zeta_{l\gamma}
     \hat{c}^\dagger_\alpha \hat{c}_j 
     \hat{c}_\delta \hat{c}^\dagger_l +
  \sum_{jkl} \zeta^*_{j\beta}\zeta_{l\gamma}\zeta_{k\delta} 
     \hat{c}^\dagger_\alpha \hat{c}_j
     \hat{c}^\dagger_k \hat{c}^\dagger_l +\nonumber\\
  &&\sum_{i}\zeta^*_{i\alpha}
     \hat{c}_i \hat{c}^\dagger_\beta 
      \hat{c}_\delta \hat{c}_\gamma+
  \sum_{ik} \zeta^*_{i\alpha} \zeta_{k\delta}
     \hat{c}_i \hat{c}^\dagger_\beta
     \hat{c}^\dagger_k \hat{c}_\gamma +
  \sum_{il} \zeta^*_{i\alpha}\zeta_{l\gamma}
     \hat{c}_i \hat{c}^\dagger_\beta
     \hat{c}_\delta \hat{c}^\dagger_l +
  \sum_{ikl} \zeta^*_{i\alpha}\zeta_{l\gamma}\zeta_{k\delta}  
      \hat{c}_i \hat{c}^\dagger_\beta
      \hat{c}^\dagger_k \hat{c}^\dagger_l  +\nonumber\\
  &&\sum_{ij}  \zeta^*_{i\alpha} \zeta^*_{j\beta}   
      \hat{c}_i \hat{c}_j
      \hat{c}_\delta \hat{c}_\gamma +
  \sum_{ijk}  \zeta^*_{i\alpha} \zeta^*_{j\beta} \zeta_{k\delta} 
      \hat{c}_i \hat{c}_j
      \hat{c}_\gamma \hat{c}^\dagger_k +
  \sum_{ijl}  \zeta^*_{i\alpha} \zeta^*_{j\beta} \zeta_{l\gamma}
       \hat{c}_i \hat{c}_j 
       \hat{c}_\delta \hat{c}^\dagger_l  + \nonumber\\
  &&\left.\sum_{ijkl}  \zeta^*_{i\alpha} \zeta^*_{j\beta} 
                \zeta_{k\delta} \zeta_{l\gamma}    
        \hat{c}_i \hat{c}_j
        \hat{c}^\dagger_k \hat{c}^\dagger_l \right] \,.
\end{eqnarray}
With the same argument needed to obtain Eq. (\ref{eq:app4}) we can write the
expectation value of the operator defined by Eq. (\ref{eq:app6}) over the
vacuum state $|0\rangle$ as  
\begin{eqnarray}
    \label{eq:app7} 
   &&\langle0|\hat{\Omega}^\dagger \hat{H}_I \hat{\Omega} |0\rangle= 
    \sum_{\alpha,\beta,\gamma,\delta} 
      U_{\alpha,\beta,\gamma,\delta} 
     \left(
       \sum_{i,j,k,l}
      \zeta^*_{i\alpha}\zeta^*_{j\beta}\zeta_{k\delta}\zeta_{l\gamma}
         \langle0| \hat{c}_i \hat{c}_j\hat{c}^\dagger_k
   \hat{c}^\dagger_l|0\rangle + \right. \nonumber \\
   &&\hspace{5.5cm}+\left.
        \sum_{i,j}\zeta^*_{i\alpha}\zeta_{j\gamma}\langle0|\hat{c}_i
        \hat{c}^\dagger_\beta\hat{c}_\delta\hat{c}^\dagger_j|0 \rangle  
     \right)
\end{eqnarray}
which is the expression given by Eq. (\ref{eq:cs3.1}).
\def\newblock{\hskip .11em plus .33em minus .07em}
%\bibliography{biblio}
%\bibliographystyle{unsrt}
 
\end{document}